  \documentclass[aps,preprint,preprintnumbers, amsmath, amssymb, prb,showpacs]{revtex4-1}
\usepackage{times}
\usepackage{graphicx}
\usepackage{psfrag}
\usepackage{ae}
\usepackage{amsmath,amssymb}
\usepackage[usenames]{color}

\begin{document}
\date{\today}

\author{Leandro B. Krott} 
\email{leandro.krott@ufrgs.br}
\affiliation{Programa de P\'os-Gradua\c c\~ao em F\'{\i}sica, Instituto 
de F\'{\i}sica, Universidade Federal
do Rio Grande do Sul\\ Caixa Postal 15051, CEP 91501-970, 
Porto Alegre, RS, Brazil}

\author{Jos\'e Rafael Bordin} 
\email{bordin@if.ufrgs.br}
\affiliation{Programa de P\'os-Gradua\c c\~ao em F\'{\i}sica, Instituto 
de F\'{\i}sica, Universidade Federal
do Rio Grande do Sul\\ Caixa Postal 15051, CEP 91501-970, 
Porto Alegre, RS, Brazil}

\title{Dynamical and structural properties of a core-softened fluid confined between fluctuating and fixed walls}

\begin{abstract}
We study the influence of mobility of the confining media in the structural and dynamical properties
of a core-softened fluid under confinement. The fluid is modeled using a two-length scale potential, which reproduces
in bulk the anomalous behavior observed in water. We perform simulations in the $NVT$ ensemble with
fixed flat walls and in the $NpT$ ensemble using a fluctuating wall control of pressure to study how 
the fluid behavior is affected by fixed and non-fixed walls. Our results indicate that the fluid 
dynamical and structural properties are strongly affected by the wall mobility. The distinct observed behavior are explained in the
framework of the two length scale potential.

\end{abstract}


\maketitle
\section{Introduction}

Water is a unique and interesting liquid. Despite abundant and common in our daily life, liquid water exhibits
several anomalous behaviors~\cite{URL}. Likewise, the properties of water attached to some substrates
or confined inside nanostructures are also of great interest for scientists, and have received an increasing attention in the last years
~\cite{MicroNano05,Bellissent04}. Understanding the behavior of water in confined geometries is essential for the comprehension of 
biological process essential to life. In addition, water plays a relevant role in new technologies based in carbon nanotubes and graphene sheets, 
as separation of fluid mixtures, water desalination and nanotubes with the capacity to mimic biological nanopores~\cite{Holt06,Shen07,Zhang08,Elimelech11, Hilder11}.
Another interesting phenomena is the shift in the region of phase transition and anomalies in the pressure temperature phase diagram induced by confinement.
Due this effect, experiments of confined water in nanotubes and nanopores have been used to avoid the spontaneous crystallization of 
water and to observe its hypothetical second critical point~\cite{Liu05,Mallamace10,Chen06, Lombardo08, Stanley11}.

Despite its molecular simplicity, there is no theoretical model able to describe all the properties of water.
The temperature and pressure dependent hydrogen bonds, the polarizability and the non-symmetrical charge distribution 
increases the task to obtain a perfect model. Hence, more than 25 distinct models based in empirical potentials
simulation were already proposed and used in all-atom Molecular Dynamics (MD) to understand
the water behavior, each one of them giving different values for the physical and chemical properties of water~\cite{URL}.
Nevertheless, several simulational works had been done in the last years using these models to understand 
the structural, dynamic and thermodynamic behavior of water in nanoconfinement~\cite{Koga01,Koga02,Gallo03,Brovchenko03,
Giovambattista09,Han10,Gallo12,Strekalova12}, and is known that different models of water can lead to significantly distinct 
results~\cite{Alexiadis08_2}.

As an alternative to the classical all-atom models of water, we can simulate water-like fluids, using the 
so-called two length scale potentials. These fluids are characterized by a simple potential
model with two characteristic length scales and, despite their simplicity, exhibit in bulk the thermodynamic, 
dynamic and structural anomalies of water~\cite{Ja98, Ya05, Xu05, Oliveira06a, Oliveira06b,Barraz09, Silva10}, and predict the 
existence of the fluid-fluid critical point hypothesized by Poole and collaborators for the ST2 water model~\cite{Po92}. 
This suggests that some of the water properties attributed to its directionality
are found in simple spherically symmetric systems.

These two length scale potential fluids offer some advantages in comparison with
the classic all-atom water models. Once this core-softened (CS) fluids had a simple interaction potential
and the entire water-like molecule is considered a sphere, is possible
to simulate large systems during long times at low computational costs. Therefore, it is possible
to investigate a large region of pressures and temperatures in the phase diagram.
Also, we can understand the anomalous properties of other systems. Material like 
liquid metals, silica, silicon, graphite, T e, Ga, Bi, S and BeF2 show thermodynamic anomalies similar to
water, while silica and silicon exhibit diffusion anomaly. However, we should address that these fluids do not have
any directionality and, therefore, are not water.

Recently the behavior of core-softened water-like fluids have been investigated in confined systems, and a similar behavior to confined water
was observed. Krott and Barbosa studied the phase diagram of a two length scale potential fluid confined between
hydrophobic walls and showed how the water-like anomalies behaves under confinement~\cite{Krott13}. Bordin and co-workers
have shown that the anomalous increase in the diffusion and the flow enhancement factor observed for water confined inside 
carbon nanotubes can also be obtained with effective models~\cite{Bordin12b, Bordin13a}. Also, Strekalova {\it {et al}}
have used CS fluids to study the properties of water confined inside a porous media~\cite{Strekalova12b}.

In addition to the question about the water model, the choice of the type of wall is not obvious.
The water behavior under confinement depends on the properties of the confining surface. Water flow
inside hydrophobic carbon and silicon-carbide nanotubes~\cite{Mao02,Ackerman03,Qin11,Khademi11}
shows a different behavior than inside hydrophilic alumina channels~\cite{Lee12}. Besides the flow, the diffusion coefficient of 
confined water depends on the confining surfaces~\cite{Zangi03a, Zangi03b,Ku05b, Han08, Franzese11, Santos12,Llave12}. 
As in the case of water confined between two walls, the roughness and the separation between the walls have a important role on 
the diffusion of the fluid \cite{eral09,choudhury10}, as the kind of particle-plate interaction~\cite{giovambattista09b, kumar07} 
also presents significant influence on the thermodynamic properties of the fluid under confinement. 
Despite the large number of recent research about fluid confined inside nanopores, 
few studies address the issue of the flexibility of the walls \cite{koga97, koga98,Ackerman03} in models of confined systems.
In general, most of these works consider rigid and fixed walls, like rocks and slit nanopores. 
We can understand a fluctuating wall as a model for the diffusion and structure of
fluids near folding proteins or confined inside biological and synthetic membranes,
as the polymer electrolyte membrane (PEM) fuel cells. PEM fuel cells are a promising type of energy conversion device, 
and the presence of water in the membrane is essential to achieve a high proton conduction~\cite{Zhao10,Eastman12}.
A better comprehension of the diffusion properties of fluids inside flexible supercapacitors~\cite{Choi11} and inside flexible metal-organic 
framework materials for gas storage/separation~\cite{Salles11} are also important for these new technological applications.
Even for gas diffusion inside nanotubes, simulational works of flexibes nanotubes indicate a distinct behavior in comparison with rigid 
nanotubes due the exchange of energy between the fluid and confining media~\cite{Chen06b,Jakobtorweihen05, Mutat12}.

These distinct systems, with rigid or flexible confinement, lead to questions about the influence
of walls mobility in the properties of a confined fluid.
In this paper we address to this question using a simple model for liquid and confining media, in order to
analyze how the wall mobility affects the dynamical and structural behavior of a confined water-like fluid. 
We study the dynamical and structural behavior of the two length scale potential model proposed by de Oliveira {\it {et al}}~\cite{Oliveira06a}
when confined between two flat walls. Two distinct scenarios were explored. In the first one, the confining walls are free 
to oscillate around an equilibrium position. To achieve this, the $NpT$ model with fluctuating walls proposed by Lupkowski and
van Smol~\cite{LupSmol90} was applied. In the second scenario the walls remain fixed, as in the case of water confined between graphene sheets. 
We show how the wall mobility influences the fluid diffusion and structure, and the distinct behaviors are discussed.

The paper is organized as follow. The water-like fluid, confining wall model and 
the details of the simulational methods used in our calculations are presented in Sec.~\ref{Model}.
Next, we discuss our results for the two scenarios in Sec.~\ref{Results},
and the conclusions of this work are summarized in Sec.~\ref{Conclu}

\section{The Model and the Simulation details}
\label{Model}

\subsection{The Model}

\begin{figure}[ht]
\begin{center}
\includegraphics[width=8cm]{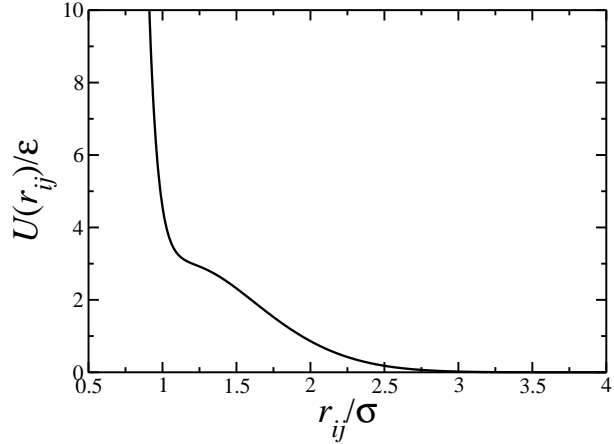}
\end{center}
\caption{Interaction potential between a water-like particles pair.}
\label{fig1}
\end{figure}

The CS fluid was modeled as point-like particles with effective diameter 
$\sigma$ and mass $m$. The fluid-fluid interaction is obtained through the three 
dimensional core-softened potential~\cite{Oliveira06a}
\begin{equation}
\frac{U(r_{ij})}{\varepsilon} = 4\left[ \left(\frac{\sigma}{r_{ij}}\right)^{12} -
\left(\frac{\sigma}{r_{ij}}\right)^6 \right] + u_0 {\rm{exp}}\left[-\frac{1}{c_0^2}\left(\frac{r_{ij}-r_0}{\sigma}\right)^2\right]
\label{AlanEq}
\end{equation}
where $r_{ij} = |\vec r_i - \vec r_j|$ is the distance between the two fluid particles $i$ and $j$.
This equation has two terms: the first one is the standard 12-6 Lennard-Jones (LJ)
potential~\cite{AllenTild} and the second one is a gaussian
centered at $r_0/\sigma$, with depth $u_0\varepsilon$ and width $c_0\sigma$.
Using the parameters $u_0 = 5.0$, $c = 1.0$ and $r_0/\sigma = 0.7$ this equation 
represents a two length scale potential, with one scale 
at  $r_{ij}\approx 1.2 \sigma$, when the 
force has a local minimum, and the other scale at  $r_{ij} \approx 2 \sigma$, where
the fraction of imaginary modes has a local minimum~\cite{Oliveira10}, as shown in Fig.~\ref{fig1}.
Despite the simplicity of the model, de Oliveira \emph{et al.}~\cite{Oliveira06a, Oliveira06b} 
showed that this fluid exhibits the thermodynamic, dynamic and 
structural anomalies present in bulk water~\cite{Kell67,Angell76}.

\begin{figure}[ht]
\begin{center}
\includegraphics[width=6cm]{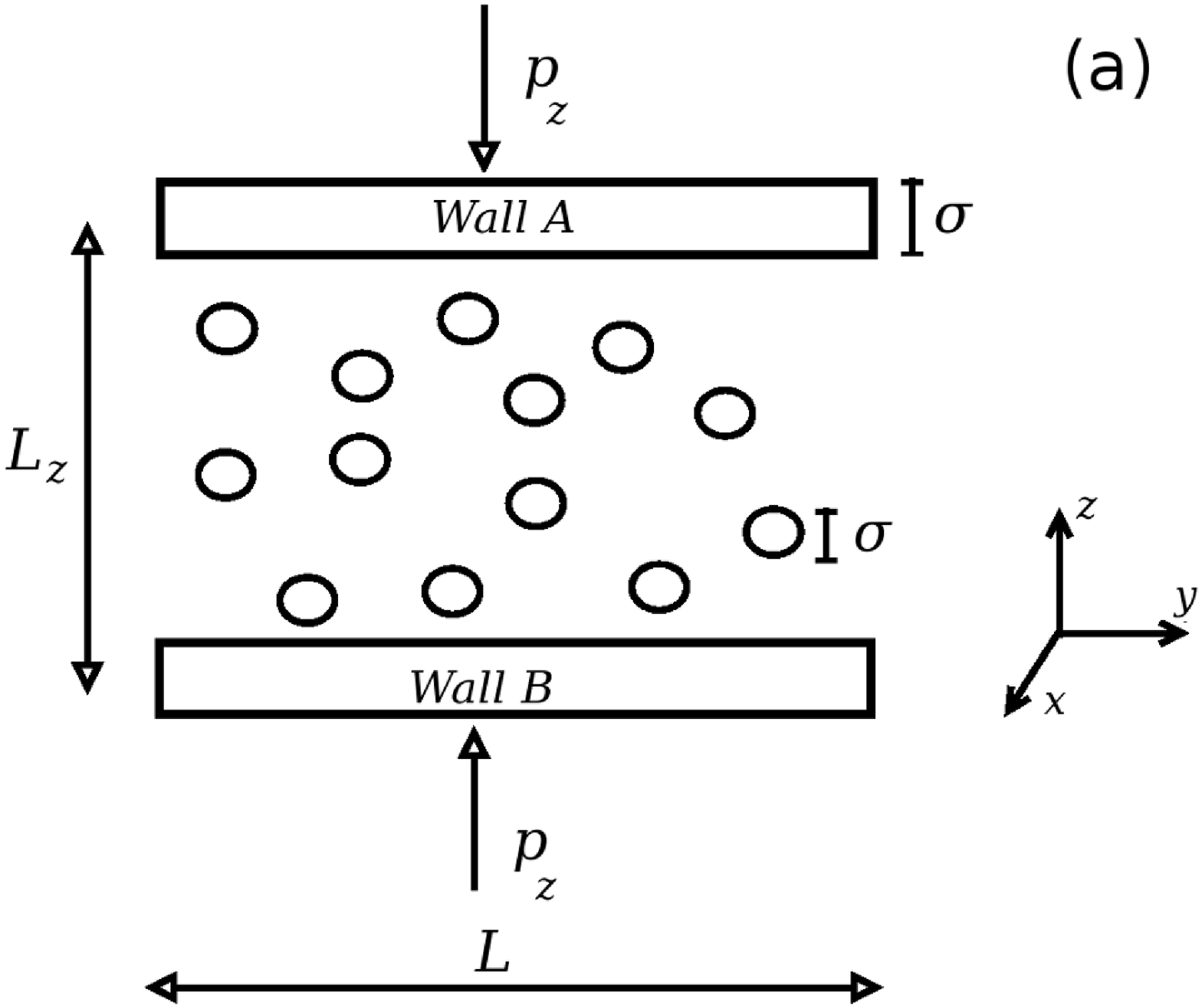}
\includegraphics[width=8cm]{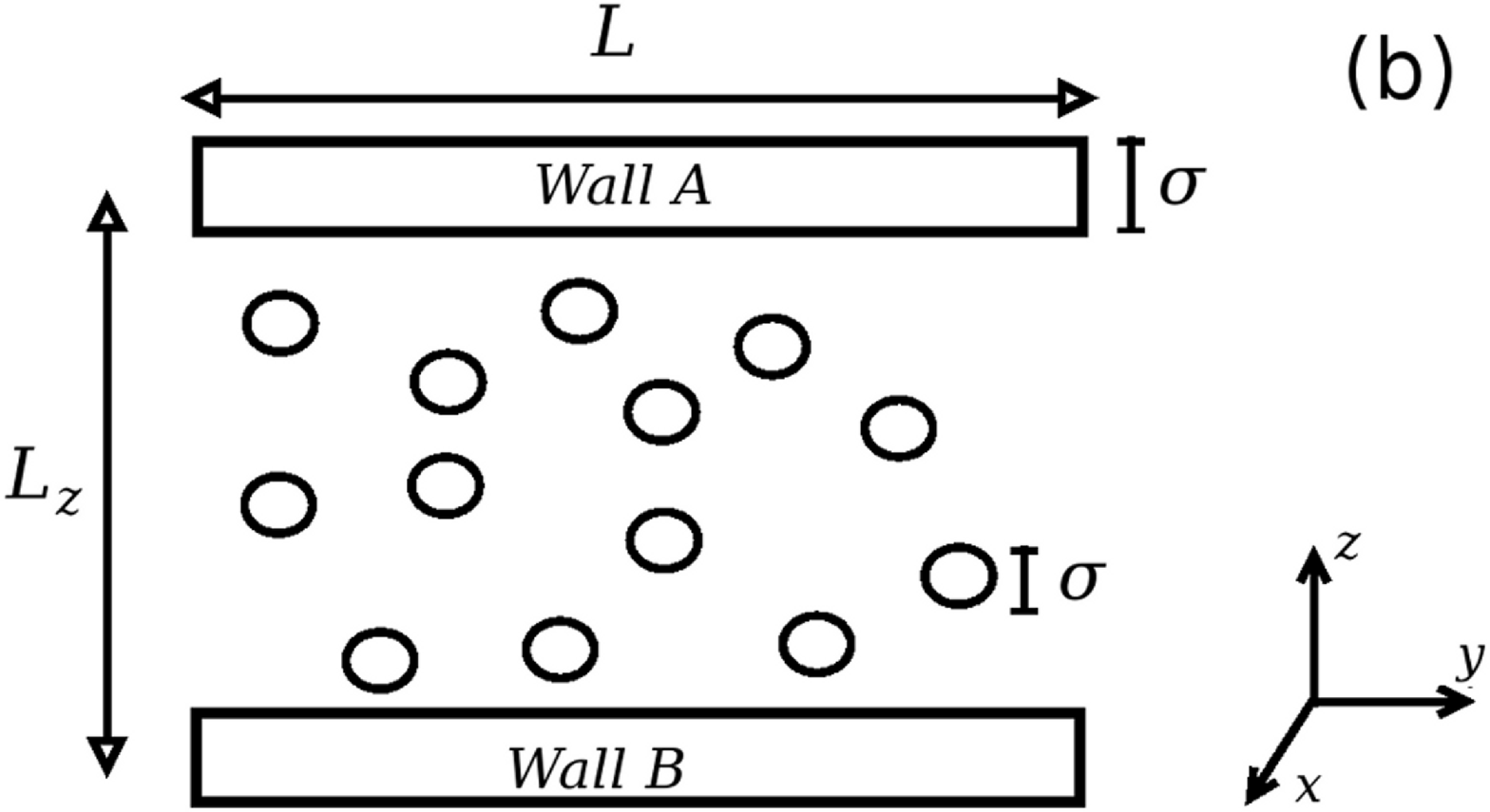}
\end{center}
\caption{Schematic depiction of the simulation box with the fluid and walls. The walls
are separated by a distance $L_z$, and have a size $\sigma$. The system with external pressure, $p_z$, and
flexible walls is shown in (a) and the system with fixed walls in (b).}
\label{fig2}
\end{figure}

Here we study the structural and dynamic behavior of this fluid confined between two parallel
plates. The simulation box is a  parallelepiped with dimensions $L_x\times L_y\times L_z$. 
The model for the fluid-wall system is illustrated in Fig.~\ref{fig2}.
Two walls, A in the top and B in the  bottom, are placed in the limits of the $z$-direction of the simulation box.
These walls are treated as fixed in the Molecular Dynamics simulations performed at constant volume, and are allowed to move
during the MD simulations realized with pressure control. The sizes $L_x$ and $L_y$ are fixed in all simulations,
and defined as $L_x = L_y = L = 20\sigma$. The values of $L_z$ were obtained first using $NpT$ simulations, 
where we have fixed the pressure $p_z$ in the $z$-direction using the Lupkowski and van Smol method of 
fluctuating confining walls~\cite{LupSmol90}, as we show in Fig.~\ref{fig2}(a). The $NVT$ ensemble simulations
was performed with separations in the same range of the $<L_z>$ obtained in the $NpT$ simulations.

The walls are flat and purely repulsive. In order to represent the interaction between a fluid particle and these walls, we
use the Weeks-Chandler-Andersen (WCA) potential. 

\begin{equation}
\label{LJCS}
U^{\rm{WCA}}(z_{ij}) = \left\{ \begin{array}{ll}
U_{{\rm {LJ}}}(z_{ij}) - U_{{\rm{LJ}}}(r_c)\;, \qquad z_{ij} \le r_c\;, \\
0\;, \qquad \qquad \qquad \qquad \quad z_{ij}  > r_c\;.
\end{array} \right.
\end{equation}
Here, $U_{{\rm {LJ}}}$ is the standard 12-6 LJ potential, included in the first term of Eq.~(\ref{AlanEq}),
and $r_c = 2^{1/6}\sigma$ is the usual cutoff for the WCA potential. Also, the term $z_{ij}$ measures the distance
between the wall $j$ position and the $z$-coordinate of the fluid particle $i$.

\subsection{The simulation details}

The properties of the system were first evaluated with MD simulations at constant number of particles, pressure and temperature 
($NpT$ ensemble). To fix the pressure in the $z$-direction, $p_{z}$, we have used the  Lupkowski and van Smol 
method~\cite{LupSmol90}. In this technique each wall has translational freedom in the confining direction
and acts like a piston in the system, where a constant force controls the pressure in the $z$-direction.

The resulting force for a water like particles was then rewrite as
\begin{equation}
 \vec F_R = -\vec\nabla U_{ij} + \vec F_{iwA}(\vec r_{iA}) + \vec F_{iwB}(\vec r_{iB})\;,
\end{equation}
where  $\vec F_{iwj}$ indicates the interaction between the 
particle $i$ and the piston $j$. 

Once the walls are allowed to move, we have to solve the equations of motion
for the pistons $A$ and $B$,

\begin{equation}
 m_w\vec a_A = p_{z}S_w\vec n_A - \sum_{i=1}^N \vec F_{iwA}(\vec r_{iA})
\end{equation}
and
\begin{equation}
 m_w\vec a_B = p_{z}S_w\vec n_B - \sum_{i=1}^N \vec F_{iwB}(\vec r_{iB})\;,
\end{equation}
respectively, where $m_w$ is the piston mass, $p_z$ the desired pressure in the 
system, $S_w$ is the piston area and $\vec n_A$ is a unitary vector 
in positive $z$-direction, while $\vec n_B$ is a negative unitary vector. Both 
pistons ($A$ and $B$) have
mass $m_w=m=1$, width $\sigma$ and area equal to $S_w = L^2$.

The system temperature was fixed using the Nose-Hoover heat-bath 
with a coupling parameter $Q = 2$. Three values of temperature was chosen:
one above the region in the phase diagram were the diffusion and the density of this fluid exhibit an anomalous behavior, 
$k_BT/\varepsilon = 0.5$, a second and a third values inside of this anomalous region, $k_BT/\varepsilon = 0.25$ and
$k_BT/\varepsilon = 0.15$~\cite{Oliveira06a}. Standard periodic boundary conditions were applied in the $x$ and $y$ 
directions. The equations of motion for the particles of the fluid were integrated using the velocity Verlet algorithm,
with a time step $\delta t = 0.005$ in LJ time units.

The fluid-fluid interaction, Eq.~(\ref{AlanEq}), has a cutoff radius $r_{\rm cut}/\sigma = 3.5$.
The number of particles was fixed in $N = 500$, and the pressure was varied from
$p_z\sigma^3/\varepsilon = 0.1$ to $p_z\sigma^3/\varepsilon = 4.5$, with a $\delta p_z\sigma^3/\varepsilon = 0.1$. 
For each pressure the systems equilibrate at different walls mean distance $<L_z>$. These displacement
was used in the $NVT$ ensemble simulations. All the others parameters are equal
to the used in the $NpT$ ensemble simulations.

Five independent runs were performed to evaluate the properties of the
confined fluid. The initial system was generated placing the fluid
particles randomly in the space between the walls. The initial displacement
for the simulations with fluctuating wall was $L_{z0} = 20\sigma$.
We performed $5\times10^5$ steps to equilibrate the system followed 
by $5\times10^6$ steps for the results production stage. The equilibration time
was taken in order to ensure that the pistons reached the equilibrium position for the fixed value of $p_z$.
For the simulations with fixed walls the particles were placed randomly in the region between the walls,
then equilibrated during $1\times10^6$ steps followed by $2\times10^6$ steps to obtain the 
physical quantities.

For calculating the lateral diffusion coefficient, $D_{||}$, we computed 
the mean square displacement (MSD) from Einstein relation
\begin{equation}
\label{r2}
\langle [r_{||}(t) - r_{||}(t_0)]^2 \rangle =\langle \Delta r_{||}(t)^2 \rangle=4Dt^\alpha\;,
\end{equation}
where $r_{||}(t_0) = (x(t_0)^2 + y(t_0)^2)^{1/2} $ and  $r_{||}(t) = (x(t)^2 + y(t)^2)^{1/2} $
denote the parallel
coordinate of the confined water-like molecule 
at a time $t_0$ and at a later time $t$, respectively.
The diffusion coefficient $D_{||}$ is then obtained from
\begin{equation}
 D_{||} = \lim_{t \to \infty} \frac{\langle \Delta r_{||}(t)^2 \rangle}{4t^\alpha}\/.
 \label{difeq}
\end{equation}
Depending on the scaling law
between $\Delta r_{||}^2$ and $t$ in 
the limit $t \rightarrow \infty$, different diffusion mechanisms can be identified:
$\alpha=0.5$ identifies a single file regime~\cite{Farimani11}, 
$\alpha=1.0$ stands for a Fickian diffusion whereas $\alpha=2.0$ refers to a 
ballistic diffusion~\cite{Striolo06,Zheng12}.
 
\section{Results and Discussion}
\label{Results}
         
\subsection*{Case A: Fluctuating walls}

\begin{figure}[ht]
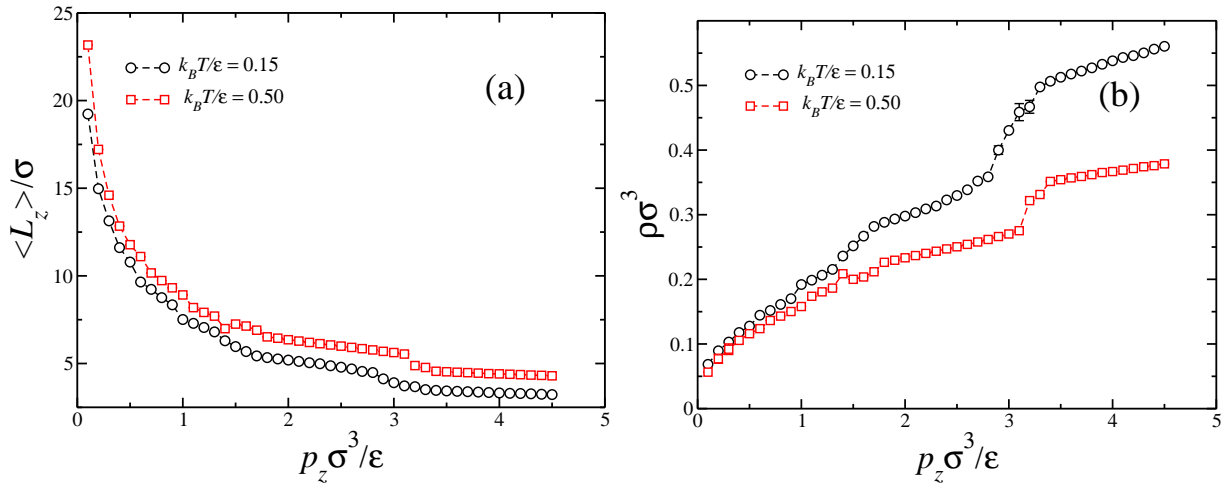

\begin{center}
\includegraphics[width=8cm]{Fig3A.eps}
\includegraphics[width=8cm]{Fig3B.eps}
\end{center}
\caption{The distance $<L_z>$ between the walls in (a) and the density $\rho$ in (b) as function
of the pressure applied by the walls in the $z$-direction for different values of temperature. 
Error bars are the deviation from the mean value. The results for $k_BT/\varepsilon = 0.25$ are not shown
here since they present the same behavior of $k_BT/\varepsilon = 0.50$.}
\label{fig3}
\end{figure}

In a first moment we analyze the behavior of the water-like fluid when confined between two fluctuating
walls. In the Fig.~\ref{fig3} we show how the separation $L_z$ between the walls in (a) and the fluid density $\rho$
in (b) vary with the applied pressure by the pistons. Since $\rho$ depends on $L_z$, both quantities exhibit the same
behavior, with small jumps at certain values of $p_z$. The small error bars indicate that the walls perform
short oscillations around the equilibrium position. In fact, for systems at $k_BT/\varepsilon = 0.25$ 
and $0.50$ $, <L_z>$ do not exhibit any error bar larger than the data point, while for $k_BT/\varepsilon = 0.15$ 
the density shows higher error bars in the region $p_z\sigma^3/\varepsilon \cong 3$. Also, the systems 
is more affected by the pressure for the smallest temperature. We can understand this as consequence 
of the competition between the external pressure pushing the walls and the opposite force generated by the fluid particles collisions
with the wall at mean velocity $<v_z> = 0.5k_BT$. To small values of temperature the pressure will act with a stronger intensity, 
while for higher values of $T$ the external force will be easily compensated by the particles collisions.
Also, the jumps observed in the density behavior can be related to structural changes
in the confined fluid. As we increase the pressure and decrease the space available for the fluid, their conformation
and structure shift, leading to a abrupt change in the numbers of layers between the walls and to the observed jumps.

\begin{figure}[ht]
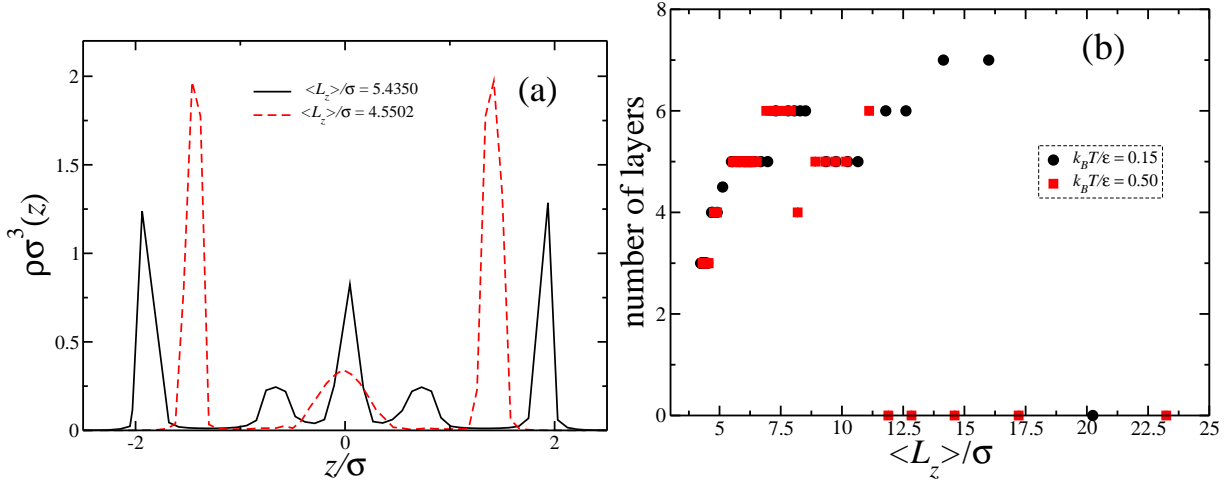

\begin{center}
\includegraphics[width=8cm]{Fig4A.eps}
\includegraphics[width=8cm]{Fig4B.eps}
\end{center}
\caption{(a) Density histograms in the $z$-direction for some values of mean separation between the walls
for systems at temperature $k_BT/\varepsilon = 0.15$ and fluctuating walls. For simplicity, we do not show the histograms for the other temperatures. 
(b) Number of layers as function of $L_z/\sigma$ for different values of temperature. Zero layers indicates
a fluid-like state.}
\label{fig4}
\end{figure}

To illustrate the different layering observed, we show in Fig.~\ref{fig4}(a) the density histogram in the confined direction 
for two values of wall displacement at temperature $k_BT/\varepsilon = 0.15$. To obtain a better comparison 
the histograms were normalized, such that 
\begin{equation}
 \int \rho(z)\quad dz = 1\;.
\end{equation}
For large values of $<L_z>$, not shown here, the fluid structurates near the walls, 
as expected for hydrophobic surfaces, but in the center a liquid-like behavior is observed.
When the system shows this structure, we assume that there is no layer formation.
As we increase the pressure and decrease the separation between the walls we observe the formation of layers, as for  
$<L_z>/\sigma = 5.4350$ and $<L_z>/\sigma = 4.5502$, where the system exhibits 5 and 3 layers, respectively. 
The histogram for all the values of temperatures and walls displacement are not shown, but the behavior
is similar. Since the temperature influences the layer formation, a distinct number of layers is observed for each value
of $T$. The dependence of the number of layers with the distance between the walls for $k_BT/\varepsilon = 0.15$ and $0.50$ is 
shown in Fig.~\ref{fig4}(b). As expected, the fluid form layers at smaller values of $<L_z>$ for higher temperatures.
The number of layers for $k_BT/\varepsilon = 0.25$ is a intermediary case between the higher and lower temperature, and
is not shown for simplicity.

The results obtained in our previous works for this water-like fluid confined between walls or inside nanotubes~\cite{Krott13, 
Bordin12b, Bordin13a} show that, when the confining structure is rigid, the fluids assumes a structure where the mean distance
between the layers in the confined direction is approximately 2. This is the characteristic distance of the second scale of the
potential Eq.~(\ref{AlanEq}). However, in our simulations with fluctuating walls the distance between the layers can be 1, which
is the characteristic distance of the first scale of the fluid-fluid interaction potential, or even smaller distances, as 
we can see in Fig.~\ref{fig4}(a) for the cases with $<L_z>/\sigma = 5.4350$ and $<L_z>/\sigma = 4.5502$. 
This unusual behavior leads the system to present a non-monotonic behavior for the number of layers as function of $<L_z>$, 
Fig.~\ref{fig4}(b), with an increase from 4 layers to 6 or 7 layers when $<L_z>/\sigma \cong 7.5$.
To understand this effect we have to remember that the layer formation is resulted of the competition between the fluid-fluid interaction, Eq.~(\ref{AlanEq}),
and the fluid-wall interaction, Eq.~(\ref{LJCS}). As consequence, we can identify regimes where the system had a high organization,
with the particles located at a distance equal to the second characteristic scale, and the enthalpic effects dominate over the 
entropic contribution for the free energy of the system. Also, the kinetic energy loss due the fluctuating walls  
favor the enthalpy, allowing the layers to accommodate at distances smaller than the second scale. 
As consequence, the system only exhibits a abrupt transitions from one number of layers to another, when the entropic contribution 
from the fluid-wall repulsion dominates. This lead to a smoother and continuous behavior of the self-diffusion coefficient
as function of the plates separation, distinct to  the one observed for this fluid when confined inside rigid 
nanotubes~\cite{Bordin12b, Bordin13a} or inside fixed walls.

\begin{figure}[ht]
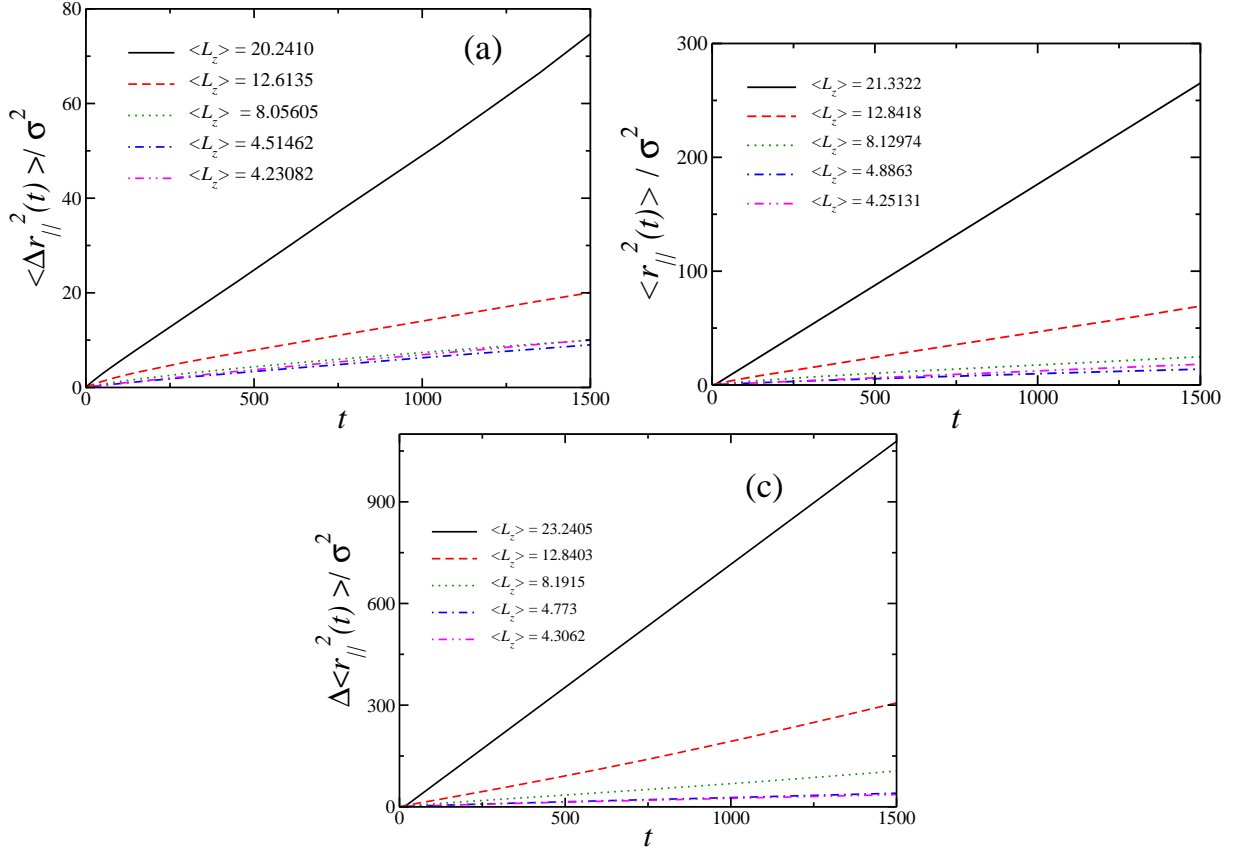

\begin{center}
\includegraphics[width=8cm]{Fig5A.eps}
\includegraphics[width=8cm]{Fig5B.eps}
\includegraphics[width=8cm]{Fig5C.eps}
\end{center}
\caption{Parallel mean square displacement versus time for some values of wall displacement and
for different values of temperature: (a) $k_BT/\varepsilon = 0.15$, (b) $k_BT/\varepsilon = 0.25$ and (c) $k_BT/\varepsilon = 0.50$
and fluctuating walls. The errors bars not shown are smaller than the symbol size.}
\label{fig5}
\end{figure}

\begin{figure}[ht]
\begin{center}
\includegraphics[width=8cm]{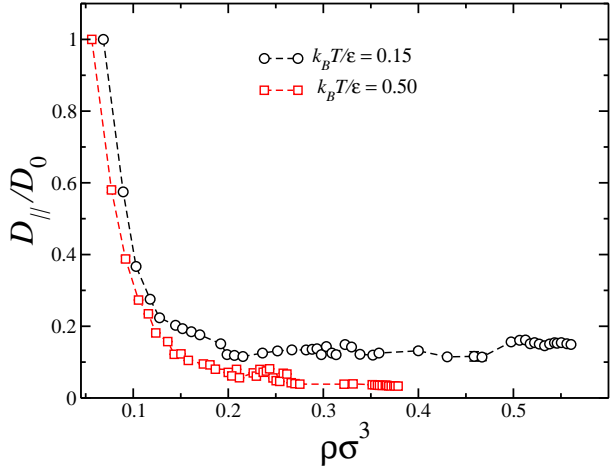}
\end{center}
\caption{ Dependence of the diffusion coefficient $D_{||}$ with the fluid density $\rho$ for systems with fluctuating walls. The system
at $k_BT/\epsilon = 0.25$ presented the same behavior of $k_BT/\epsilon = 0.50$.}
\label{fig6}
\end{figure}

In order to understand the dynamical properties of this system we evaluate
the MSD $\Delta r_{||}^2$ to check the diffusive regime of the fluid. In our model, we found $\alpha = 1$, or a Fickian diffusion,
for all plates displacement and for all temperatures, as shown in Fig.~\ref{fig5}. The diffusion coefficient $D_{||}$ was then
evaluated using the Eq.~(\ref{difeq}). We show the behavior of $D_{||}$ as function of the fluid density 
in Fig.~\ref{fig6}. To maintain the graphics for different temperatures in the same scale we define $D_0$ as the value
of the self diffusion coefficient at the smallest pressure for each isotherm.
Our results indicates a rapid decrease in $D_{||}$ as we increase $\rho$ (or $p_z$), and then a saturation.
The saturation occurs when the fluid assumes a layering structure and, after this, $D_{||}$ exhibits 
small fluctuations, that are related to changes in the number of layers of the confined fluid. 
This rapid saturation and the continuous and smooth shape of the curve for $D_{||}$ reinforce our assumption that the fluctuating walls
favor the fluid to assume a structure that increases enthalpy and decreases the entropy, decreasing also the mobility.
How the isotherms indicate, the temperature seems do not play a relevant role in this case, since the qualitative behavior is the
same for all values of $k_BT/\epsilon$. 

Is equally important to notice the absence of an anomalous behavior in the diffusion, characterized by a
region where the diffusion coefficient increases as the density of the system increases.
The diffusion anomaly was observed for all-atom models of water confined between
hydrophobic fixed plates~\cite{Zangi03b} and inside nanotubes~\cite{Farimani11, Zheng12}, and even for our water-like
fluid model confined inside nanotubes~\cite{Bordin12b}. Our results indicates that allowing the walls to fluctuate, even
when the fluctuation is small, lead to complete distinct results for the dynamical and structural behavior. 

\subsection*{Case B: Fixed walls}

Here we analyze systems confined between fixed walls, whose values of $L_z/\sigma$ were chosen in the same range of the mean values obtained for 
the fluctuating cases discussed in the last section. 

\begin{figure}[!htb]
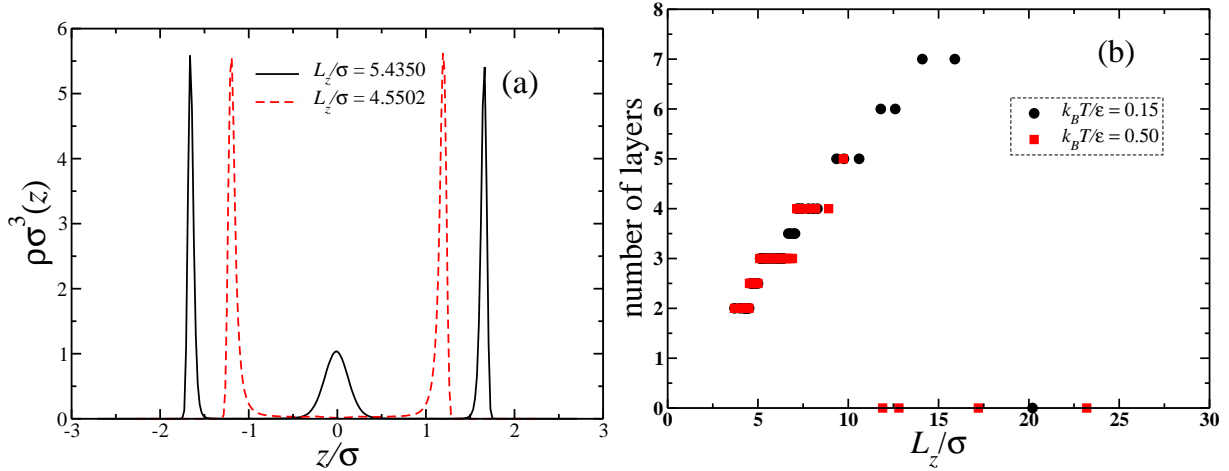

\begin{center}
\includegraphics[width=8cm]{Fig7A.eps}
\includegraphics[width=8cm]{Fig7B.eps}
\end{center}
\caption{(a) Density histograms in the z-direction for some values of separation between the walls for systems at $k_BT/\varepsilon = 0.15$ and fixed walls. 
(b) Number of layers as function of $L_z/\sigma$. Zero layers indicate a bulk-like behavior.}
\label{fig7}
\end{figure}

The formation of layers is shown in the Fig.\ref{fig7}(a) through the density histograms for some values 
of $L_z/\sigma$, at $k_BT/\varepsilon = 0.15$. Comparing these cases with those presented in the Fig.\ref{fig4}(a), 
we verified that the structuration of the particles between the fixed walls are very different in relation to 
fluctuating cases at same mean values of $L_z/\sigma$.
For $L_z/\sigma = 5.4350$ and $4.5502$, the systems with fixed walls presented the formation of 3 and 2 layers, respectively, while for 
fluctuating cases, 5 and 3 layers were formed. The structuration of the fluid for both kinds of confinement is similar
just for high distances between the walls.

\begin{figure}[ht]
\begin{center}
\includegraphics[width=8cm]{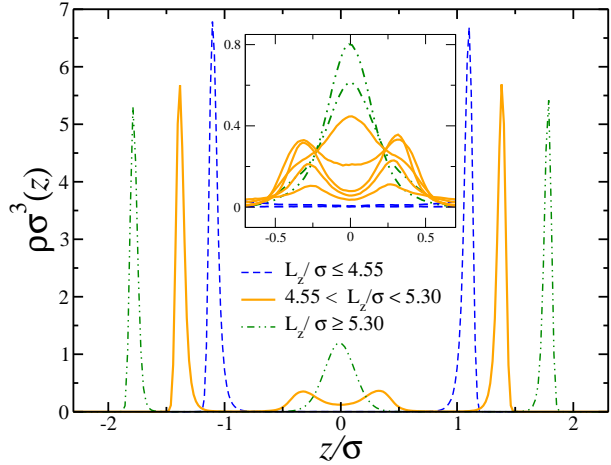}
\end{center}
\caption{Density histograms showing the two-to-three layers transition at $k_BT/\varepsilon = 0.15$ and fixed walls. 
The inset shows a zoom of the middle layer, where the formation of sublayers occurs.}
\label{fig8}
\end{figure}

The Fig.\ref{fig7}(b) shows the number of layers formed for $k_BT/\varepsilon = 0.15$ and $0.50$. The systems at $k_BT/\varepsilon = 0.25$
are not shown for simplicity. Bulk-like systems occur for walls separated by high distances and they are indicated by 
zero layers. Broken numbers of layers, given by $2.5$ and $3.5$, indicate systems with formation 
of sublayers. A system with $2.5$ layers, for example, means the formation of two well defined contact layers and one 
middle layer with sublayers. When the systems change of a number of layers to another, presenting 
sublayers at intermediate cases, we called of layering transition. A two-to-three layers systems at $k_BT/\varepsilon = 0.15$ are shown
in the Fig.\ref{fig8}. The same behavior is observed for $k_BT/\varepsilon = 0.25$ and $0.50$. Since the walls are fixed, there is a limitation for 
the structuration of the particles between the walls. This favors the entropy and, as consequence, there will be the formation of sublayers. In counterpart, 
the fluctuation of the walls allows the accommodation of the particles in smaller distances, like the first scale of the fluid 
potential ($r/\sigma \sim 1$), as explained in last section. A transition between layers also was observed in systems
like the SPC/E model confined between hydrophobic plates \cite{Giovambattista09, B805361H}.

\begin{figure}[ht]
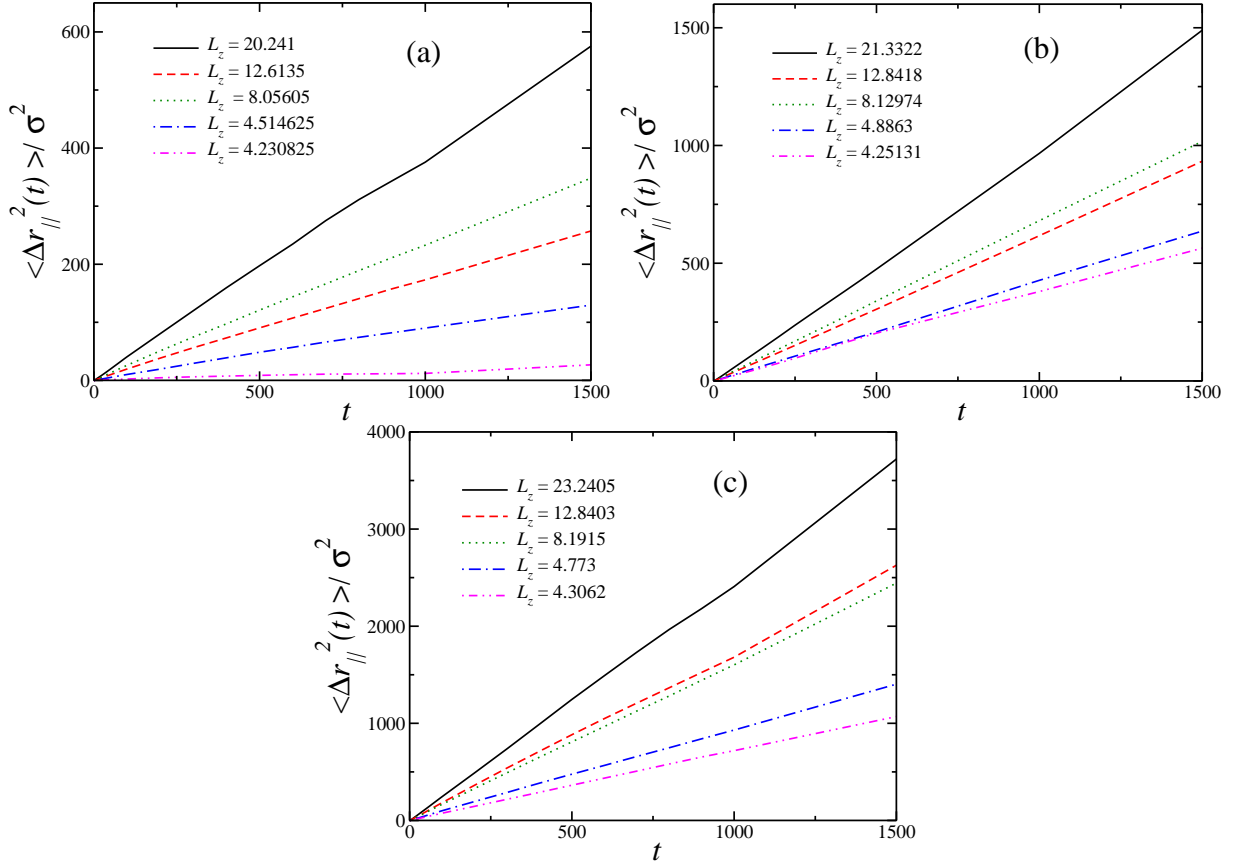

\begin{center}
\includegraphics[width=8cm]{Fig9A.eps}
\includegraphics[width=8cm]{Fig9B.eps}
\includegraphics[width=8cm]{Fig9C.eps}
\end{center}
\caption{Mean square displacement for different separations between the plates at $k_BT/\varepsilon = 0.15$ in (a), $k_BT/\varepsilon = 0.25$ in (b) and
$k_BT/\varepsilon = 0.50$ in (c) in systems with fixed walls.}
\label{fig9}
\end{figure}

Besides the layering density, we also are interested to understand what happens with the dynamic of the systems when the walls are fixed.
We show in Fig.\ref{fig9} the parallel mean square displacement ($\Delta r_{||}^2$)
as function of time, for some values of $L_z/\sigma$ at $k_BT/\varepsilon = 0.15$ in (a),  $0.25$
in (b) and  $0.50$ in (c). Like seen for fluctuating walls, the systems with fixed walls presented a Fickian diffusion regime, where $\alpha = 1$.
Considering that, the diffusion coefficient was calculated using the Eq.~(\ref{difeq}).

\begin{figure}[ht]
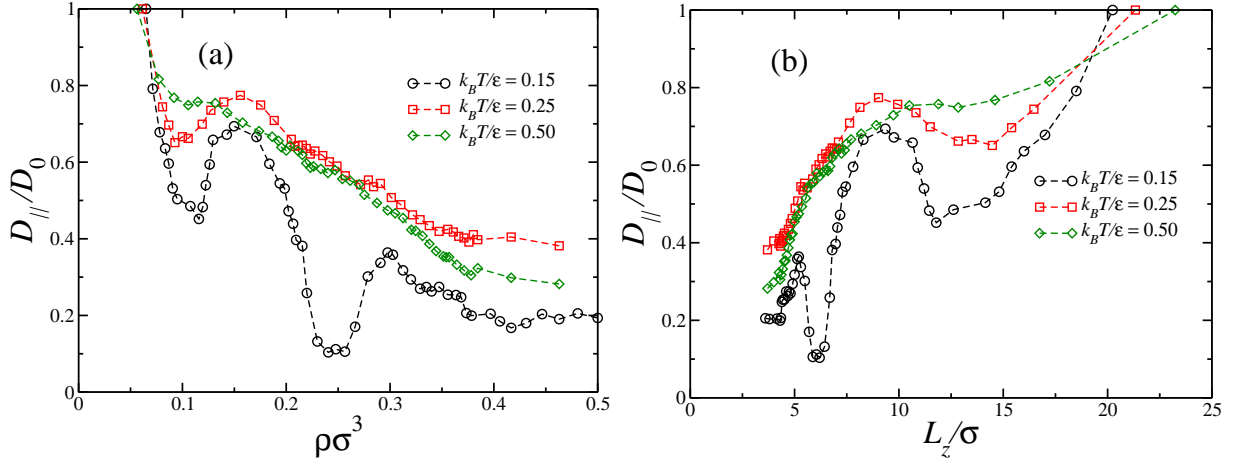

\begin{center}
\includegraphics[width=8cm]{Fig10A.eps}
\includegraphics[width=8cm]{Fig10B.eps}
\end{center}
\caption{Dependence of the diffusion coefficient $D_{||}$ with fluid density $\rho$ in (a) and the distance $L_z/\sigma$ in (b) for systems with fixed walls.}
\label{fig10}
\end{figure}

With the intuition to obtain a better understanding of the dynamical properties of the systems, we show in Fig.\ref{fig10}(a) 
the diffusion coefficient normalized by $D_0$ as function of density.  At low densities,
ranged between $\rho \sigma^3 \approx 0.100$ and $0.150$, an anomalous behavior in its diffusion is observed for the
isotherms $k_BT/\varepsilon = 0.15$ and $0.25$. At high temperatures, like $k_BT/\varepsilon = 0.50$, this anomalous behavior 
almost disappears. The anomaly in the diffusion for this model already was observed in the bulk system~\cite{Oliveira06a},
in a previous work for confining rough fixed plates~\cite{Krott13} and inside nanotubes~\cite{Bordin12b}. As we saw in the last section in Fig.\ref{fig6} 
(a) and (b), the anomalous behavior in the diffusion did not appear for fluctuating walls. 

\begin{figure}[ht]
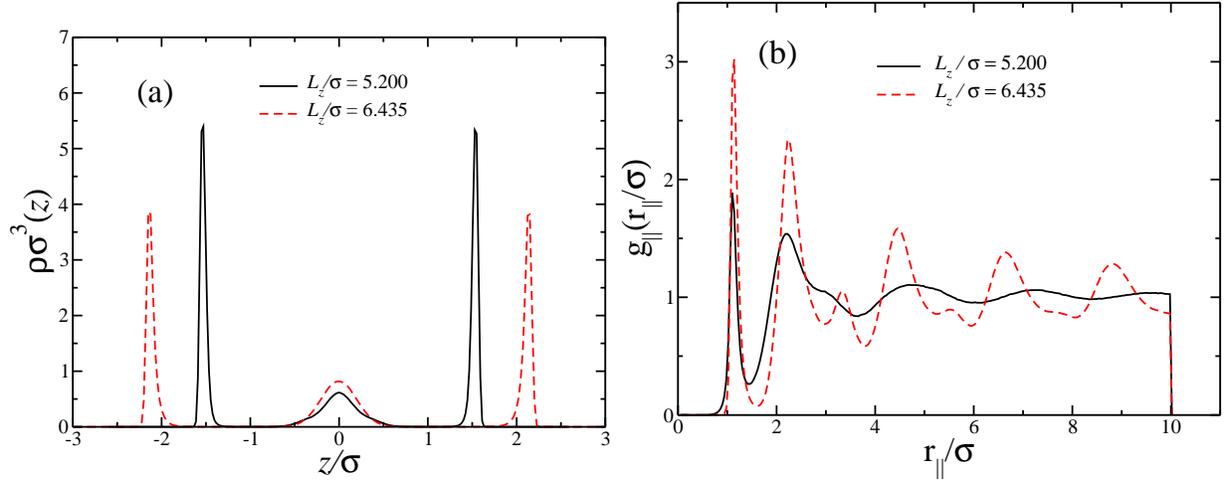

\begin{center}
\includegraphics[width=8cm]{Fig11B.eps}
\includegraphics[width=8cm]{Fig11C.eps}
\end{center}
\caption{Case B: In (a), we show the density histograms for $L_z/\sigma = 5.200$ and $6.435$
and in (b), we have the radial distribution function for 
the middle layer of $L_z/\sigma = 5.200$ and $6.435$. 
}
\label{fig11}
\end{figure}

Addictionally to the diffusion anomaly, we can observe an amorphous phase for some densities in the isotherm $k_BT/\varepsilon = 0.15$. To clarify this behavior,
we show in Fig.\ref{fig10}(b) the diffusion coefficient as function of $L_z/\sigma$, and it is possible to see the values of $L_z/\sigma$ where the 
diffusion has a significant decrease, characterizing an amorphous behavior of the system. As example,
we consider a system at $L_z/\sigma = 5.200$, characterized by a liquid-like behavior, and other at $L_z/\sigma = 6.435$, 
corresponding to an amorphous system. Both systems present the formation of three well defined layers, like we
can see in the density histograms, in Fig.\ref{fig11}(a). The structure of the system can be analyzed using 
the radial distribution function $g_{||}(r_{||})$ for each layer, defined as

\begin{equation}
\label{gr_lateral}
g_{\parallel}(r_{\parallel}) \equiv \frac{1}{\rho ^2V}
\sum_{i\neq j} \delta (r-r_{ij}) \left [ \theta\left( \left|z_i-z_j\right| 
\right) - \theta\left(\left|z_i-z_j\right|-\delta z\right) \right].
\end{equation}

\noindent where the Heaviside function $\theta (x)$ restricts the sum of particle pair in a slab of thickness $\delta z = 1.0$.

For systems with three well defined layers, we can calculate two $g_{||}(r_{||})$, one to each layer. The $g_{||}(r_{||})$ for 
the middle layer of the systems at $L_z/\sigma = 5.200$ and $6.435$ is shown in Fig.\ref{fig11}(b). The smooth behavior of
the $g_{||}(r_{||})$ for $L_z/\sigma = 5.200$ indicates a liquid-like structure, while the sharp peaks for $L_z/\sigma = 6.435$
indicate a amorphous structure of the system. The $g_{||}(r_{||})$
for the contact layers are not shown because the conclusion about the structuration of the particles is the same. 
The systems at $L_z/\sigma = 5.872$, $6.047$ and $6.197$ also presented amorphous behavior in their diffusion and 
structure. The solidification for some distances $L_z$ is supported by many works, using 
TIP5P~\cite{Zangi03b, Ku05b} and SCP/E~\cite{Giovambattista09} models.

\section{Conclusion}
\label{Conclu}

We have studied the structural and dynamical behavior of a core-softened fluid confined between two walls.
This fluid, in bulk, exhibits the thermodynamic, dynamic and structural anomalies observed in liquid water.
We have consider two kinds of confinement. In the first, the walls are allowed to fluctuate, while in the second case
the walls are fixed. In both systems the fluid-wall interaction is by excluded volume purely, and the wall is a simple planar plate.
Our simulations shows that if we consider a system were the confining walls are free to move, fluctuating around
the equilibrium position, the structure and the self diffusion of the fluid are different from the obtained for
the case with fixed walls. The small fluctuations of the wall increase the entropic contribution to the free energy
of the fluid, leading to structural conformations not observed when the confining media is fixed. Also, the diffusion
coefficient diffusion shows a smooth dependence with the wall displacement, saturating when the system exhibits a layering.
For fixed plates, besides the transition between layers and the diffusion anomaly, we observe a amorphous behavior 
for some values of $L_z$ at low temperature, $k_BT/\varepsilon = 0.15$, what is not observed for the fluctuating cases.
For small temperatures, the self-diffusion coefficient exhibits minimals, relates to this amorphous structure.

\section{Acknowledgments}

This work was partially supported by the CNPq and CAPES.

 \bibliographystyle{aip}

\end{document}